\documentclass[aps,pre,twocolumn,float]{revtex4}
\usepackage{amsmath,bm,epsfig}
\let \nn  \nonumber
\usepackage{pstricks}
\usepackage{pst-node}
\usepackage[ansinew]{inputenc}
\usepackage{amssymb,amsmath}
\usepackage{multirow}

\def\p{\partial}

\def\a{\alpha}

\def\e{\varepsilon}

\def\o{\omega}
\def\wt{\widetilde}

\def\be{\begin{equation}}       \def\ba{\begin{array}}

\def\ee{\end{equation}}         \def\ea{\end{array}}

\def\bea {\begin{eqnarray}}      \def\eea {\end{eqnarray}}

\def\bean{\begin{eqnarray*}}    \def\eean{\end{eqnarray*}}

\def\RA {\ \Rightarrow\ }

\newtheorem{exi}{Example}

\begin{document}

\title{Time scales and structures of wave interaction exemplified with water waves}
\author{Elena Kartashova}
 \email{Elena.Kartaschova@jku.at}
  \affiliation{$^{*}$Institute for Analysis, J. Kepler University, Linz, Austria}

   \begin{abstract}
Presently two models for computing energy spectra in weakly nonlinear dispersive media are known: kinetic wave turbulence theory, using a statistical description of an energy cascade over a continuous spectrum (K-cascade), and the D-model,  describing resonant clusters and energy cascades (D-cascade) in a deterministic way as interaction of distinct modes.
In this Letter we give an overview of these structures and their properties and a list of criteria, which model of an energy cascade should be used in the analysis of a given experiment, using water waves as an example. Applying time scale analysis to weakly nonlinear wave systems modeled by the focusing nonlinear Sch\"{o}dinger equation, we demonstrate that K-cascade and D-cascade are not competing processes but rather two processes taking place at different time scales, at different characteristic levels of nonlinearity and based on different physical mechanisms.
Applying those criteria to data known from experiments with surface water waves we find, that the energy cascades observed occurs at short characteristic times compatible only with a D-cascade.  The only pre-requisite for a D-cascade being a focusing nonlinear Sch\"{o}dinger equation, the same analysis may be applied to existing experiments with wave systems appearing in hydrodynamics, nonlinear optics, electrodynamics, plasma, convection theory, etc.
\end{abstract}

{PACS: 47.27.Ak, 52.25.Fi, 05.45.-a}

\maketitle
\section{Introduction}
Wave interaction theory  describes the interaction of waves in terms of energy exchange among weakly nonlinear modes.
Two main types of energy transport  have been intensively studied: exchange of energy among a small  number of modes, for example an isolated resonant triad or quartet, and time evolution of the wave field as a whole consisting of an infinite number of modes.

\emph{Interaction of distinct modes.} The dynamic equations for a resonant triad and for a resonant quartet have been solved analytically, \cite{whitt,SS05,Cr85}; dynamics and kinematics of independently evolving resonance clusters formed by isolated or connected resonant triads or quartets is studied in  \cite{CUP}.
The increment chain equation method (ICEM)  for computing the energy spectrum of a cascade formed by distinct modes (further referred to as D-cascade) was introduced in \cite{K12a}. In \cite{K12b}, the D-model of weakly nonlinear wave interaction was presented, which integrates in a single theoretical frame  D-cascades and resonance clusters.

\emph{Interaction of an infinite number of modes.} The first oceanographic studies aiming to describe wave field evolution of surface water waves in terms of  energy cascades  date back to the beginning of the twentieth century. A major breakthrough was achieved by Hasselmann in 1962, in \cite{has1962}, giving a statistical description of an energy cascade in terms of a wave kinetic equation.  The stationary solution of this kinetic equation (further referred to as K-cascade)  for capillary waves has been found in 1967 by Zakharov and Filonenko, \cite{zak2}, coining the term  "weak turbulence". The method was extended to other classes of dispersive weakly nonlinear wave systems and renamed to wave turbulence theory (WTT), in \cite{ZLF92}.

Experimental studies performed through the last decades  revealed numerous phenomena such as dependence of the form of energy spectra on the excitation parameters \cite{NR11}, exponential form of the energy spectrum \cite{Erik}, nonlinear frequencies of the modes forming a cascade \cite{SWW-09}, etc., which could not be covered by or even were in contradiction to the  WTT.  It is also to be said, that in some cases the temporal or spatial scales required for the predictions of WTT to occur make it impractical to check them in any real experiment. For example,
in  the recent review of Newell and Rumpf \cite{NR11}, the estimate is given that the formation of a K-cascade in surface water waves with wavelength of 60 meters  "would require a tank of approximately 60 km" in length.  A tank of 60 km in length  would have to be in the open air, and each gust of wind would destroy the assumptions of the kinetic WTT about the wave fields being close to stationarity and being homogeneous in space.

These problems were addressed by the invention of new models of WTT, e.g. \cite{Pu99,zak4,K06-1,LPPR09}, all of them relying on the statistical model of an energy cascade in Fourier space (with resonances of linear modes as underlying physical mechanism).

The paradigm changed with the advent of the D-cascade which uses a deterministic model for an energy cascade (with modulation instability as underlying physical mechanism).
Presently an experimentalist has the choice between two models for an energy cascade  - K-cascade and D-cascade - to describe an experiment.
In this Letter we give criteria for this choice  and apply them to various types of water waves as examples.

We demonstrate that a K-cascade and a D-cascade are not competing processes, but rather two processes, which  take place at  different time scales, at different characteristic levels of nonlinearity and which are based on different physical mechanisms - resonant interactions of linear modes (K-cascade)  and modulation instability (D-cascade).
Also the form of the cascade is different: for the dispersion function $\o(k) \sim k^{\a}, \a\neq 1,$
a K-cascade is decaying according to \emph{power law}  and its form is independent of the total energy in the wave system. A D-cascade is decaying \emph{exponentially} and its form depends on the energy
contained in the wave system.

Taking surface water waves as an example, we show, that energy cascades in this system  occur at much faster characteristic times than those required by the kinetic WTT but can  be described as  D-cascades.

 At the end of this Letter, a list of important properties is given which can be used to determine from a given set of experimental or numerical data
 which type of cascade - D-cascade or K-cascade or even both - should be used to describe these data.
\section{Time scales}
Regard a nonlinear dispersive PDE of the form
\bea
 D \, \psi(x,t)= F(\psi, \psi_x, \psi_t, \psi_{xx}, \psi_{xt}, \psi_{tt},..) \,  \, \mbox{where} \label{general}\\
 D= \frac{\p}{\p_t}+\sum_{j=0}^{J}(-1)^jA_j\frac{\p^{2j+1}}{\p_{x^{2j+1}}},\\
D \, \varphi(x,t)=0 \RA \varphi(x,t)=A \exp[i(kx-\o(k)t)],\\
 \o^2(k)=\sum_{j=0}^{J}A_jk^{2j+1} >0, \, v(k)=\frac{d \o(k)}{d k} \neq \mbox{const}. \label{conditions}
\eea
Conditions (\ref{conditions}) provide that the dispersion function $\o(k)$ is a real function and not linear in $k$.

For studying (\ref{general}) with small nonlinearity we follow \cite{Cal92} introducing a small parameter $0<\e \ll 1$ and assuming that
the solutions of (\ref{general}) have again the form $\wt{A} \exp[i(kx-\o(k)t)]$, but with slowly changing amplitudes $\wt{A}(t\varepsilon^m,x\varepsilon^m)$. Then the nonlinearity $F$ can be rewritten as
\bea
F(\e\psi, \e\psi_x, \e\psi_t, \e\psi_{xx}, \e\psi_{xt}, \e\psi_{tt},..)=  \nn \\
\sum_{m=2}^{M}\e^mF^{(m)}(\psi, \psi_x, \psi_t, \psi_{xx}, \psi_{xt}, \psi_{tt},..)+\mathcal{O}(\e^{M+1}). \label{1-esp}
\eea
The use of any multi-scale method (in essence a transformation of variables)  then allows to study the properties of a given solution to (\ref{general}) at each time scale $\sim 1/\e^{m}$ separately.
Depending on the  chosen level of nonlinearity, different physical phenomena -- resonance clusters, D-cascade and K-cascade -- can occur as it will be shown in the Sec.3.

\section{Structures}
\emph{{\textbf{3.1. D-cascade, $\e\sim 10^{-1}$.}}}\\
Beginning  with
one wave, any multi-scale method inevitably yields -- at time scale $1/\e^2$  --  the nonlinear  Schr\"{o}dinger equation (NLS)  in "slow" variables $T, X$:
\be
i\a_1\wt{A}_T+\a_2\wt{A}_{XX}+\a_3|\wt{A}|^2\wt{A}=0, \label{NLS}
\ee
(for simplicity written in 1+1 dimensions, for details see e.g.  \cite{Cal92}).  The slowly changing amplitude $A(X,T)$  of the wave depends on slow variables $T=\e^2 \, t, \, X=\e^2 \, x$; with  $x$ and $t$  being real physical space and time. 

An  NLS is called focusing if $\a_3/a_2>0$,  and therefore admits modulation instability which the physical mechanism underlying the formation of a D-cascade.
In this section we restrict ourselves to the case of a focusing NLS.

Modulation instability is the effect that, under certain conditions,  a wave train $\o_0$, called carrier wave,  becomes modulated by two side-band modes with slightly different frequencies $\o_1$ and $\o_2$ fulfilling conditions  $\o_1 + \o_2 = 2\o_0, \ \
{\bf k}_1+{\bf k}_2=2{\bf k}_0,$ where
$\o_1=\o_0 + \Delta \o, \, \o_2=\o_0 - \Delta \o, \, 0<\Delta \o \ll 1$, \cite{BF67}. For each mode $\o_0$ its instability interval can be written and the most unstable mode within this interval can be computed, also for modified NLS-s, e.g. \cite{DY79,Hog85}. The most unstable modes are called cascading modes while they form a D-cascade, \cite{K12a}.

For computing the frequencies and amplitudes of the cascading modes,  at each cascade step $n$ the increment chain equation method (ICEM) is used, \cite{K12a}, yielding
two chain equations
\bea
 \o_{n+1}=\o_n +  \o_n A(\o_n) k_n, \label{chain-plus}\\
    \o_{-(n+1)}=\o_{-n} -  \o_{-n} A(\o_{-n}) k_{-n}, \label{chain-minus}
\eea
describing the formation of a unidirectional D-cascade, for direct and inverse cascades respectively.  The direct and inverse cascade are not symmetrical, \cite{KSh11}. The chain equations may be used to compute the form of the energy spectrum, the direction of a cascade, conditions of cascade termination, etc., \cite{K12b}. The form of D-cascade is exponential as $A^2_n=p^nA_0^2, \, 0<p<1$ with some constant $p$.
After $n$ cascade steps, $n$ cascading and $(2^n-n)$ non-cascading modes are excited; non-cascading modes provide spectrum broadening.

It follows from  (\ref{chain-plus}),(\ref{chain-minus}) that the frequencies of the cascading modes are nonlinear, i.e. dependent on amplitudes. It can be shown that (\ref{chain-plus}),(\ref{chain-minus}) describe exact resonances of nonlinear Stokes waves in Zakharov's equation.  If the nonlinearity gets too small,  all modes are linear and a D-cascade is not formed. Instead, under some additional conditions \cite{Has67}, resonance clusters formed by linear frequencies may occur.

\emph{Remark 1.} For a D-cascade described by (\ref{chain-plus}),(\ref{chain-minus}), the characteristic level of nonlinearity is $\e\sim 0.1\div 0.25$, and the  chain equation is computed from the NLS. For bigger nonlinearity, say $\e\sim 0.25\div 0.4$, a modified NLS should be taken with nonlinear terms up to order 4; for details see \cite{DY79,Hog85}. The corresponding chain equations are given in \cite{K12a}.

\textbf{\emph{{3.2. Resonance clusters, $\e\sim 10^{-2}$.}}}\\
In a wave system in which  resonance of  $N$ waves is possible, $ N\ge 3$, the application of a multi-scale method yields a system of equations of the form (\ref{NLS}) with interconnected coefficients whose solutions satisfy resonance conditions $\sum_{j=1}^{N}\pm\o({\bf k}_j)=0, \, \sum_{j=1}^{N}\pm {\bf k}_j=0$. Neglecting the "slow" space variables, one gets the well-known dynamical systems for 3- and 4-wave resonances, respectively,  of the form
\bea
i\dot{B}_1=  V^3_{12} B_2^*B_3,\,
i\dot{B}_2=  V^3_{12} B_1^* B_3, \, i\dot{B}_3=  - V^3_{12} B_1 B_2;\label{3-wr}\\
\begin{cases}\label{4-wr}
  i\, \dot{B}_1=  V^{12}_{34} B_2^*B_3B_4 , \,,
 i\, \dot{B}_2=V^{12}_{34} B_1^*B_3B_4,  \\
 i\, \dot{B}_3=  (V^{12}_{34})^* B_4^*B_1B_2, \,,
 i\, \dot{B}_4=  (V^{12}_{34})^* B_3^*B_1B_2
 \end{cases}
\eea
in canonical variables $B_j$. Like the slowly changing amplitudes $A_j$ defined in the previous section, the canonical variables $B_j$ are functions of "slow" time: in the 3-wave system $B_j=B_j(\e \, t)$ and in the 4-wave system  $B_j=B_j(\e^2 \, t)$. By suitable transformation of the variables $A_j$, any Hamiltonian system may be transformed into a form as given by (\ref{3-wr}) and (\ref{4-wr}), respectively. This form is called "canonical form", the corresponding transformation "canonical transformation". The canonical form of a dynamical system is the simplest possible form of the dynamical system in the respect, that dynamical systems written out in  canonical variables have the same form for different physical systems; the difference between the physical systems is completely hidden in the interaction coefficient specific to the dynamic system -  $ V^3_{12}$ for a 3-wave systems and $V^{12}_{34}$ for a 4-wave system.

The systems (\ref{3-wr}),(\ref{4-wr}) describe the \emph{primary} resonance clusters in 3- and 4-wave systems. In these wave systems also independent \emph{common} clusters may be formed which consist of primary clusters having joint modes; they form the bridge to distributed initial state necessary for K-cascade. The form of common clusters can be computed by  $q$-class decomposition, \cite{KK06-1,KK06-2,KK06-3}.

Time-scales for 3- and 4-wave resonances are $T_{3-res}\sim 1/\e$ and $T_{4-res}\sim 1/\e^2$, respectively. They  are called dynamic time scales.

\textbf{\emph{{3.3. K-cascade, $\e\sim 10^{-2}$.}}}\\
An initial energy distribution over  \emph{an infinite number of resonance clusters} is the prerequisite of kinetic WTT, \cite{ZLF92}.
In order to go to the kinetic regime, one has to make assumptions (including, but not restricted to: existence of an infinite number of primary resonance clusters with linear frequencies interconnected in a special way, a wave field close to the stationary state, homogeneity in space, existence of an inertial interval where energy is conserved) which are necessary for the  mathematical deduction of the wave kinetic equation.

By statistical ensemble averaging, dynamics of individual modes (at the dynamic time scale) is averaged out, and field evolution is described at the much longer kinetic time scale.
The 3-wave kinetic equation reads as
\bea
\frac{\bf d }{{\bf d } t}{B}^2_3=
\int |V^3_{12}|^2
\delta(\o_3-\o_1-\o_2)\delta({\bf k}_3-{\bf k}_1-{\bf k}_2) \nn \\
\cdot (B_1B_2-B_1^{*}B_3-B_2^{*}B_3)
{\bf d}{\bf k}_1 {\bf d}{\bf k}_2 \label{3w-KE}
\eea
and is valid at the  time scale $T_{3-kin} \sim 1/\e^2$. For a 4-wave system a similar kinetic equation is deduced  at the time scale $T_{4-kin} \sim 1/\e^4$ and so on for any $s$-wave system with finite $s$. Time scales $T_{3-kin}, T_{4-kin},...$ are called kinetic time scales. An $N$-wave K-cascade is a stationary solution of the $N$-wave kinetic equation and occurs at the kinetic time scale $T_{N-kin} \sim 1/\e^{2(N-2)}$. K-cascade decays according to power law $k^{\a}$ where $\a$ is constant for a given
  linear dispersion relation and does not depend on the total energy in a wave system.

The fact that for any $N$-wave system the kinetic time scale is substantially longer than the resonance time scale can be understood in the following way: a wave system  needs a long time to move from independently evolving \emph{finite resonance clusters} via non-resonant interactions to a homogeneous and almost stationary state, with energy distributed over  \emph{an infinite number of modes}.

\emph{Remark 2.} Both for resonance clusters and K-cascade, the small parameter $\e$ must be small enough to exclude nonlinear frequency shift;  in water wave systems it is usually taken as $\e \sim 0.01$ .
If the nonlinearity gets bigger, the dispersion relation becomes dependent on amplitudes yielding a nonlinear frequency shift, and neither q-class decomposition nor kinetic equation method can be used. No general methods are known for describing resonance clusters or K-cascades for nonlinear frequencies.

  We now can answer the question, which effect, starting from zero initial conditions, will be observed first in an experiment with water wave interactions. The answer will depend mainly on the level of nonlinearity.

  Small level of nonlinearity ($\e \sim 0.01$): If the geometry of the experimental tank and (related to it) the dispersion function permit three wave resonances or near resonances, we will see 3-wave resonant clusters appearing at time scale $\sim \e^{-1}$. Kinetic WTT for this case predicts a K-cascade to take place at time scale $\sim \e^{-2}$, which means, that anything we are going to see at time scale $\sim \e^{-1}$ must be a resonant cluster.

  If there are no 3-wave interactions, again depending on the shape of the experimental tank and the dispersion function, 4-wave interactions have to be taken into account, leading to 4-wave resonant clusters taking place at time scale $\sim \e^{-2}$. The corresponding K-cascade by design has time scale $\sim \e^{-4}$, so again anything observed at time scale $\sim \e^{-2}$ has to be a resonant 4 wave cluster.

   Moderate level of nonlinearity ($\e \sim 0.1 \div  0.4$): independent of dispersion function and shape of experimental tank we are going to see a D-cascade at time scale $\sim \e^{-2}$, which is in real time, due to the much higher value of $\e$, comparable to or even faster than the time for a 3 wave resonant cluster.
   We illustrate this below, computing the characteristic time for both cascades and for various types of water waves.

\begin{table*}
\begin{tabular}{|c|c|c|c||c|c||c||c|c|}
\hline
 wave length & wave type &dispersion relation& wave period&$T_{3-res}$ &$T_{3-kin}$& $T_{MI}$ & $T_{4-res}$&$T_{4-kin}$\\
\hline
1 mm & capillary&$\o^2=\frac{\sigma}{\rho} k^3$&0,0022 sec &0,22 sec&22 sec&0,22 sec&-&-\\
\hline
10 cm &gravity-capillary&$\o^2=\frac{\sigma}{\rho} k^3 +gk$&0,25 sec&25 sec& 42 min& 25 sec &-&-\\
\hline
1 m &surface gravity& $\o^2=gk$ &0,8 sec&-&-&80 sec &2,5 hours& 2,63 years\\
\hline
10 m &surface gravity&$\o^2=gk$ &2,53 sec&-&-&4 min &7 hours& 8 years\\
\hline
\end{tabular}
\caption{\label{t:2} Characteristic times for formation of D-cascade, K-cascade and resonance clustering in various water waves (at 20°C): the gravitational
constant of acceleration $g=9,8m/sec^2$; density $\rho=10^3kg/m^3$; the coefficient
of surface tension $\sigma=72,75\cdot10^{-3}kg\cdot m/sec^2$. Characteristic levels of nonlinearity are taken as following: $\e_{MI} \sim 10^{-1}, \, \e_{res} \sim 10^{-2}$}
\end{table*}
\section{Characteristic interaction times for water waves}

Let us first regard surface water waves with $\o^2=gk$ and  wave length $\lambda=1 m$. Then the wave number $k=2\pi/\lambda\approx  6,3 m^{-1}$, the wave frequency $\o=(9,81 \cdot 6,3)^{1/2} \approx 7,86 sec^{-1} $ and the wave period is $t=2\pi/\o \approx 0.8 sec$. Corresponding characteristic times are:

for a D-cascade: $\e\sim 10^{-1} \RA T_{MI}\sim t/\e^2=t/10^{-2} \sim $ 80 sec,

for 4-wave resonances: $\e\sim 10^{-2} \RA T_{4-res}\sim t/\e^2=t/10^{-4} \sim $ 8000 sec $\approx$ 2,5 hours,

for a 4-wave K-cascade: $\e\sim 10^{-2} \RA T_{4-kin}\sim t/\e^4=t/10^{-8} \sim$ about 3 years.

 In
Table \ref{t:2} examples  are given of the characteristic time, this is the time, measured in wave periods, it takes for a wave phenomenon to occur  for different types of water waves and typical wave lengths. Water waves have been chosen as an example because at least some,  though not many, measurements of the time scale of wave field evolution are known. For instance, it is established experimentally that the "fast" wave field evolution of gravity surface waves (caused by a sudden wind increase) occurs after only a few dozen wave
periods \cite{Aut95}. The difference between time scales, say, $T_{MI}$ and $T_{4-kin}$ is also very well known to physicists working on present-day wave forecasting systems, e.g. \cite{PJ04}.

\section{Summary and conclusions}
In this Letter we have shown that in the focusing NLS three structures of wave interaction may occur - D-cascade, resonance clusters and K-cascade. Their properties are listed below.

\emph{D-cascade}:\\
1D) nonlinearity $ \e_{MI} \sim 10^{-1}$;\\
2D)  generated by modulation instability;\\
3D) built of nonlinear frequencies;\\
4D) occurs at the dynamic time scale $T_{MI} \sim t/\e_{MI}^2$;\\
5D) decays exponentially; its form depends on the total energy in the wave system.

\emph{K-cascade}:\\
1K) nonlinearity  $\e_{res} \sim 10^{-2}$;\\
2K)   generated by $N-$wave resonant interactions;\\
3K)  built of linear frequencies;\\
4K)  occurs at  kinetic time scales, for instance, $T_{3-kin} \sim t/\e_{res}^2$ for 3-wave interaction and $T_{4-kin} \sim t/\e_{res}^4$  for 4-wave interaction;\\
5K) decays according to a power law; its form  does not depend on the total energy in the wave system.

\emph{Resonance cluster}:\\
1R) nonlinearity $ \e_{res} \sim 10^{-2}$;\\
2R)  generated by $N-$wave resonant interactions;\\
3R)  built of linear frequencies;\\
4R)  occurs at  dynamic time scales, for instance, $T_{3-res} \sim t/\e_{res}$ for 3-wave interaction and $T_{4-res} \sim t/\e_{res}^2$  for 4-wave interaction.

For any set of data, either from experiment or from direct numerical simulation with evolution equations, \emph{all of the criteria} given above must be checked in order to decide which model - D-cascade or K-cascade - should be used to describe them.

As the D-cascade is a very young concept, one can not expect to find experimental studies interpreted in terms of a D-cascade. But, as D-model has no special pre-requisites, it may be rewarding to re-evaluate existing experiments, especially if they use high levels of nonlinearity  or are even known to have an exponential energy spectrum, as in \cite{XiSP10,Erik} (capillary water waves).

Moreover, as it was shown in previous section, all experiments with surface water waves and other 4-wave systems, e.g. \cite{FFF08,BOBC08,Mor10}, should be re-evaluated while K-cascade is not observable in real time in these systems.

A D-cascade as discussed in this Letter relies on the physical mechanism of modulation instability which is found "in numerous physical situations including
water waves, plasma waves, laser beams, and electromagnetic transmission lines" \cite{ZO09}. An  application area of special interest is  the formation of extreme waves  for which
modulation instability plays a central role, e.g. \cite{freak-theor3,MO-all09,SS11,SDP11,DNP13}.

\textbf{{Acknowledgments.}}   The author is grateful to the Referee for most useful remarks which allowed to improve the presentation.
The author is very much obliged
to the organizing committee of the workshop ``Integrable systems -- continuous and discrete -- and transition to chaos" (Centro Internacional de Ciencias, Cuernavaca,
Mexico; November-December 2012), where part of this work was accomplished. The author is specially grateful to  F. Calogero, E. Pelinovsky, P. Santini and H. Tobisch for fruitful discussions. This work has been  supported by the Austrian Science Foundation (FWF) under project	 P22943.

\end{document}